\documentclass[a4paper]{article}
\usepackage{INTERSPEECH2022,amsmath,amssymb,graphicx}
\usepackage[bookmarks=false,pdftex,pdfpagemode=None,pdfstartview=FitH,pdfview=FitH,pdftitle={End-to-end LPCNet: A Neural Vocoder With Fully-Differentiable LPC Estimation},pdfauthor={Krishna Subramani, Jean-Marc Valin, Umut Isik, Paris Smaragdis, Arvindh Krishnaswamy}]{hyperref}
\usepackage{balance}

\title{End-to-end LPCNet: A Neural Vocoder With Fully-Differentiable LPC Estimation}

\name{Krishna Subramani\sthanks{\ \ Work performed while at Amazon Web Services}\ $^{\flat}$, Jean-Marc Valin$^\natural$, Umut Isik$^\natural$, Paris Smaragdis$^{\flat\natural}$, Arvindh Krishnaswamy$^\natural$}
\address{$^\flat$ University of Illinois at Urbana-Champaign\quad$^\natural$ Amazon Web Services}
\email{ks51@illinois.edu, \{jmvalin, umutisik, parsmara, arvindhk\}@amazon.com}

\begin{document}
\maketitle

\begin{abstract}
Neural vocoders have recently demonstrated high quality speech synthesis, but
typically require a high computational complexity.
LPCNet was proposed as a way to reduce the complexity of neural
synthesis by using linear prediction~(LP) to  assist an autoregressive model.
At inference time, LPCNet relies on the LP coefficients being explicitly computed from
the input acoustic features.
That makes the design of LPCNet-based systems more complicated, while
adding the constraint that the input features must represent a clean speech
spectrum. 
We propose an end-to-end version of LPCNet that lifts these limitations
by learning to infer the LP coefficients from the input features in the frame rate network .
Results show that the proposed end-to-end approach equals or exceeds the
quality of the original LPCNet model, but without explicit LP analysis.
Our open-source\footnote{Source code available at
\url{https://github.com/xiph/LPCNet/} in the \texttt{lpcnet\_end2end} branch.}
end-to-end model still benefits from LPCNet's low complexity, while allowing
for any type of conditioning features.
\end{abstract}
\noindent\textbf{Index Terms}: neural speech synthesis, end-to-end optimization, linear prediction, LPCNet, WaveRNN
\section{Introduction}
\label{sec:intro}

Vocoder algorithms are used to synthesize intelligible speech from acoustic
parameters.
Over the past few years, vocoders based on deep neural networks (DNN) have vastly exceeded
the capabilities of classical techniques such as linear predictive
vocoders~\cite{markel1974linear} and
concatenative~\cite{hunt1996unit} synthesizers.
This has resulted in significant quality
improvements for text-to-speech~(TTS) synthesis, speech compression, and other
speech processing applications.


Many different types of neural vocoders have been proposed in the past, including
autoregressive models such as WaveNet~\cite{oord2016wavenet} and
WaveRNN~\cite{kalchbrenner2018efficient}, but also GANs~\cite{donahue2019wavegan},
and flow-based algorithms~\cite{prenger2019waveglow}. Most
of these algorithms require either a GPU or a powerful CPU to synthesize
speech in real time, limiting their use to servers or powerful devices. 
The LPCNet~\cite{valin2019lpcnet} vocoder, an improvement over WaveRNN, was
recently introduced as a way of reducing the computational cost of neural synthesis.
It uses linear prediction (LP) and the source-filter model of speech
production~\cite{fant1960acoustic} to simplify the task of the DNN model.
The combination of signal processing with DNNs makes it possible to synthesize
high quality speech with a complexity  reducing the computational cost of
$\sim\,$3 GFLOPS, making LPCNet suitable for many existing phones. 

LPCNet has so far been used in multiple applications, including 
compression~\cite{valin2019lpcnetcodec}, TTS~\cite{kons2019tts} and voice
conversion~\cite{zhao2021conversion}.
One limitation of the original LPCNet algorithm is that
it requires explicit computation of the LP coefficients from the acoustic features
at inference time. That requirement
makes it impossible to use an arbitrary latent feature space, or even acoustic features that
do not correspond to the clean speech being synthesized. LPCNet is thus
unsuitable for a range of applications that includes
end-to-end speech coding~\cite{zeghidour2021soundstream},
end-to-end TTS~\cite{donahue2021tts},
or cases where the input does not represent a full clean spectrum,
such as bandwidth extension~\cite{gupta2019bwe}.

In this work, we propose an end-to-end differentiable LPCNet
(Section~\ref{sec:lpcnet}) that avoids the explicit LPC computation.
Instead, we learn LPC estimation from the input features
by backpropagating the gradient from an improved loss function
(Section~\ref{sec:e2elpcnet}). We evaluate the end-to-end LPCNet on a
language-independent, speaker-independent task (Section~\ref{sec:evaluation}).
Results show that the proposed algorithm achieves
the same quality as the original LPCNet (Section~\ref{sec:results}) while
removing the need for the explicit LPC computation.
That makes LPCNet suitable across a wide range of
applications and devices (Section~\ref{sec:conclusion}).

\section{LPCNet Overview}
\label{sec:lpcnet}
WaveRNN~\cite{kalchbrenner2018efficient} predicts a discrete probability 
distribution function~(pdf) $P(s_t)$ for each sample $s_t$ at time $t$ from
conditioning information, and the past samples up to $s_{t-1}$. It uses a gated recurrent unit (GRU), 
followed by linear layers and a softmax activation to output the distribution. 
The GRU uses block-sparse weight matrices to reduce the computational 
requirements. 

LPCNet~\cite{valin2019lpcnet} builds upon WaveRNN, using linear
prediction to simplify the DNN's task.
LP coefficients $a_i$ are explicitly
computed from the input cepstral features, by first turning them back into a spectrum,
from which it computes the autocorrelation, followed by the Levinson-Durbin 
algorithm. The LPCs are then used to compute a prediction $p_t$ from the
previous samples:
\begin{equation}\label{eq:prediction}
p_t = \sum_{i = 1}^{M} a_{i}s_{t - i}\ ,
\end{equation}
where $M$ is the prediction order. For LPCNet operating at 16~kHz, we use $M=16$.
The excitation, or residual, is then defined as
$e_t = s_t - p_t$\ .

The main GRU ($\mathrm{GRU_A}$) can then use not only the past
signal $s_{t-1}$, but also the past excitation $e_{t-1}$ and the prediction for the
current sample $p_t$. Similarly, the output estimates the distribution of the
excitation $P(e_t)$. 

To avoid WaveRNN's two-pass coarse-fine strategy to output the full
16-bit pdf, LPCNet uses the \mbox{$\mu$-law} scale
\begin{align} \label{eq:mulaw}
U(x) = \mathrm{sgn}(x) \cdot \frac{U_{\mathrm{max}}\log(1 + \mu |x|)}{\log(1 + \mu)}\ ,
\end{align}
where the \mbox{$\mu$-law} range is $U_{\mathrm{max}}=128$ and $\mu = 255$. Although
\mbox{$\mu$-law} values are typically represented as positive in the $[0,256[$ range,
in this paper we treat them as signed values within $[-128,128[$ to simplify notation,
with $U(0)=0$.

The \mbox{$\mu$-law}
scale -- which is also used in the original WaveNet proposal -- makes the
quantization noise independent of the signal amplitude. Furthermore,
a pre-emphasis filter $E(z)=1-\alpha z^{-1}$ applied to the input
avoids audible quantization noise from the \mbox{8-bit}
\mbox{$\mu$-law} quantization. The inverse, de-emphasis filter
$D(z)=\frac{1}{1-\alpha z^{-1}}$ is applied on the synthesis output.
We use $\alpha=0.85$.

A frame rate network learns frame conditioning features $\mathbf{f}$ from the cepstral 
features. A sample rate network predicts the output probability 
distribution for the excitation $e_t$ when given the prediction $p_t$, the past 
excitation $e_{t - 1}$ and the past signal value $s_{t - 1}$, along with the 
frame conditioning feature $\mathbf{f}$. 

\begin{figure}[t]
\centering
\includegraphics[scale=0.5]{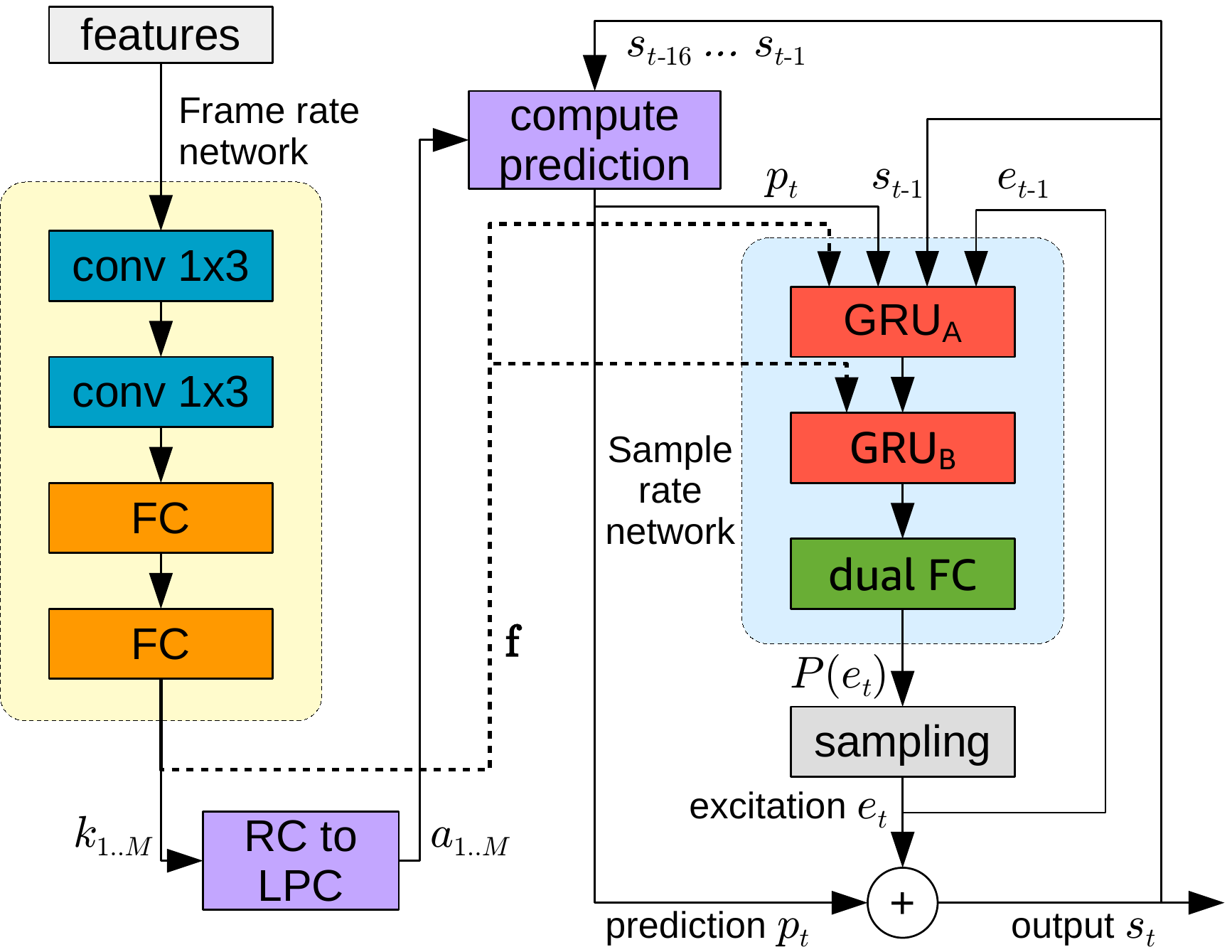}
\caption{Proposed end-to-end LPCNet architecture. The frame rate network operates
on frames and computes the conditioning features $\mathbf{f}$, including the
RCs $k_i$ to be converted to LP coefficients $a_i$.
The sample rate network predicts each sample, using the previous samples,
as well as the excitation and prediction.}
\label{fig:e2e_overview}
\end{figure}

\section{End-to-end LPCNet}
\label{sec:e2elpcnet}

We propose an end-to-end version of LPCNet where the LPC
computation is differentiable and integrated into the algorithm rather
than hardcoded based on a fixed set of cepstral features.
Doing so makes it easier to use LPCNet, in a wider range of applications.
The
overview in Fig.~\ref{fig:e2e_overview} shows that the main difference
with the original LPCNet is that the frame rate network is now used
to compute the LPCs.
To train a fully differentiable, end-to-end LPCNet model, three separate aspects
have to be made differentiable:
\begin{enumerate}
	\vspace{-0.1em}\item LPC computation,
	\vspace{-0.1em}\item input embedding, and
	\vspace{-0.1em}\item loss function computation.
	\end{enumerate} 
At inference time, only the LPC computation differs from the original
LPCNet. 

\subsection{Learning to Compute LPCs}
LP coefficients are very sensitive to error -- and can easily become
unstable -- which is why they are never directly quantized or estimated.
More robust representations include reflection
coefficients (RC)~\cite{makhoul1975linear}, which guarantee stability as long as
they lie in the $]-1, 1[$ interval, and line spectral
frequencies~\cite{itakura1975line}, which need to follow an alternating
ordering. Between those, we choose to estimate RCs,
not only because the~$]-1, 1[$ interval is easily enforced by the use of
a $\mathrm{tanh}()$ activation, but also because it can be shown that the
pre-tanh "logit" is equal (up to a scaling factor) to the
log-area ratio~\cite{viswanathan1975quant} representation of RCs, which
is known to be robust to error.

RCs can be converted to direct-form LP coefficients
with the Levinson recursion~\cite{levinson1947lpc}
\begin{equation}\label{eq:levinson}
a_j^{(j)} = 
\begin{cases}
    k_i,         & \text{if } j=i\\
    a_j^{(i-1)}+k_i a_{i-j}^{(i-1)}, & \text{otherwise}
\end{cases}
\end{equation}
where $k_i$ are the RCs and $a_j^{(i)}$ are the $j^{\text{th}}$
prediction coefficients of order $i$. Since both~(\ref{eq:prediction})
and~(\ref{eq:levinson}) are differentiable, the network can
learn to compute RCs.

We directly use the first $M$~elements of the frame rate network condition
vector~$\mathbf{f}$ (without increasing its dimensionality) as the RCs.
This not only simplifies the design, but also ensures that the
following sample rate network can take into account the estimated prediction coefficients.
The complexity of both~(\ref{eq:levinson}) and the cepstral-to-LPC conversion
being replaced is negligible compared to the total inference complexity, therefore
the proposed end-to-end system has essentially the same complexity as the original
LPCNet.

The SFNet vocoder~\cite{rao2020sfnet}
also has its feature network learn RCs
that match the spectral characteristics of the target synthesized speech.
In the case of LPCNet, we want to directly learn the LP representation that
optimizes the final loss function.

\subsection{Differentiable Embedding Layer}
The inputs $p_{t}$, $s_{t - 1}$, $e_{t-1}$ are quantized to 8-bit \mbox{$\mu$-law} values,
which are then used to learn an input sample embedding.
We find that the embedding learns a set of non-linear functions that are applied to
the input, improving synthesis quality compared to directly using the signal
values as input to the GRU.
However, $\mu$-law quantization of the prediction and excitation prevents the
gradient from backpropagating the loss all the way to the RC computation in the frame rate network.
To make the embedding differentiable, we propose a simple interpolation scheme.

Let $\mathbf{v}_j$ be the $j^\mathrm{th}$ embedding vector of the embedding matrix and
$x$ be a real-valued (unquantized) $\mu$-law sample value. We can compute the interpolated
embedding $\mathbf{v}^{(i)}\left({x}\right)$ as
\begin{align}
f &= x - \lfloor x \rfloor\\
\mathbf{v}^{(i)}\left({x}\right) &= (1 - f) \cdot \mathbf{v}_{\lfloor x \rfloor} + f \cdot \mathbf{v}_{\lfloor x \rfloor + 1}\ .
\end{align}
The approach allows a gradient to propagate through the fractional interpolation
coefficient $f$. 

At inference time, we avoid the (small) extra complexity of the interpolation and compute
$\mathbf{v}\left({x}\right) = \mathbf{v}_{\lfloor x \rceil}$, where $\lfloor \cdot \rceil$
denotes rounding to the nearest integer.

\subsection{Loss function}
In the original LPCNet, the excitation can be pre-computed before the DNN training.
In the proposed end-to-end LPCNet, the excitation is computed as part
of the DNN architecture. The ground-truth target for the cross-entropy loss is thus
no longer constant, raising two problems for our loss function. The first problem
is that -- as was the case for the embedding -- quantizing the target $\mu$-law
excitation would not allow the gradient to backpropagate to the RC computation.
We propose a similar solution to the interpolated embedding: an interpolated
cross-entropy loss. Let $e^{(\mu)}_t$ be the real-valued $\mu$-law excitation
at time $t$ and let $\hat{P}(e^{(\mu)}_t)$ be the discrete probability estimated
for time $t$. Rather than rounding to the nearest integer and using the standard
cross-entropy loss
\begin{equation}
\mathcal{L}_\mathrm{CE} = \mathbb{E}\left[ -\log \hat{P}\left(\left\lfloor e^{(\mu)}_t \right\rceil\right) \right]\ ,
\label{eq:loss-cross-entropy}
\end{equation}
we interpolate the probability such that the loss becomes
\begin{align}
f &= e^{(\mu)}_t - \left\lfloor e^{(\mu)}_t \right\rfloor\\
\hat{P}^{(i)} \left( e^{(\mu)}_t \right) &= (1-f)\hat{P}\left(\left\lfloor e^{(\mu)}_t\right\rfloor \right)
                                                  + f \hat{P}\left(\left\lfloor e^{(\mu)}_t\right\rfloor+1 \right)\\
\mathcal{L}_\mathrm{ICE} &= \mathbb{E}\left[ -\log \left( \hat{P}^{(i)} \left( e^{(\mu)}_t \right) \right)\right]\ ,
                                                  \label{eq:interpolated-crossentropy}
\end{align}
where the expectation is taken over the distribution of the data over time. Again, the interpolation coefficient $f$ helps the gradient backpropagate.

The second problem we have with our loss -- and one that still applies to
(\ref{eq:interpolated-crossentropy}) -- has to do with the non-linear nature
of the $\mu$-law scale. The $\mu$-law spacing is wider for large excitation
values and narrower for excitation values close to zero. It means that for the
same linear uncertainty, the cross-entropy is larger when the excitation is
large, \textit{i.e.} when the predictor is \textit{worse}. 

We need to minimize a cross-entropy loss corresponding to the
distribution of the linear-domain excitation samples. This could be done by
simply dividing $\hat{P}^{(i)}\left(\cdot\right)$ by the linear step size corresponding
to a difference of 1 in the $\mu$-law value. Instead, since we are already interpolating
to a continuous distribution, we divide by the derivative of the $\mu$-law expansion
function $U^{-1}(\cdot)$ to obtain a compensated loss:
\begin{equation}
  \mathcal{L_\mathrm{C}} = \mathbb{E}\left[-\log \frac{\hat{P}^{(i)} \left( e^{(\mu)}_t \right)}{\left.\frac{d}{dx}U^{-1}(x)\right|_{e^{(\mu)}_t}} \right]\ .\label{eq:compensated-loss0}
\end{equation}
Taking advantage of the fact that $U^{-1}(\cdot)$ is piecewise exponential and
discarding constant terms in the loss, (\ref{eq:compensated-loss0})~simplifies to
\begin{equation}
  \mathcal{L}_\mathrm{C} = \mathcal{L}_\mathrm{ICE} + \mathbb{E}\left[\frac{\left|e^{(\mu)}_t\right|\log (1+\mu)}{U_\mathrm{max}}\right]\ ,\label{eq:compensated-loss}
\end{equation}
where the left-hand term $\mathcal{L}_\mathrm{ICE}$ is our previously-derived
interpolated cross-entropy loss
in~(\ref{eq:interpolated-crossentropy}) and the right-hand term compensates
for the $\mu$-law scale with an $L_1$~loss on the $\mu$-law companded excitation.

\subsection{Regularization}

We observe that using the compensated loss $\mathcal{L}_\mathrm{C}$ alone often
leads to the optimization process diverging. To stabilize the training,
we add regularization to the LPC estimation. We consider the following three different
regularization variants.

\subsubsection{$L_1$ regularization}
\label{sec:L1-regularization}
Our $\mu$-law compensation in (\ref{eq:compensated-loss}) already includes an
$L_1$~loss, so the obvious regularizer is to artificially increase that compensation term using
\begin{equation}
  \mathcal{L}_{L_1} = \gamma\mathbb{E}\left[\frac{\left|e^{(\mu)}_t\right|\log (1+\mu)}{U_\mathrm{max}}\right]\ .\label{eq:compensated-loss-gamma}
\end{equation}
We consider $\gamma=1$, which effectively doubles the compensation term from
(\ref{eq:compensated-loss}) and is sufficient to stabilize the training.
In effect, the regularization minimizes the prediction error in a similar way to the
standard LPC analysis. Operating on the $\mu$-law residual has the desirable property
that the regularization applies almost equally to high- and low-amplitude signals.

\subsubsection{Log-area ratio regularization}
\label{sec:LAR-regularization}
Another way of regularizing the LPCs is to directly attempt to match the
"ground truth" LPCs used by the original LPCNet. Since those LPCs are computed on
the target speech signal and are only needed at training time,
the regularization does not impose additional constraints on the end-to-end LPCNet.  

As a distance metric between the estimated RCs, $k_i$, and the ground truth RCs, $k^{(g)}$, we use the
log-area ratio (LAR), which is both easy to compute and representative of the
difference between the two filters:
\begin{equation}
  \mathcal{L}_\mathrm{LAR} = \mathbb{E}\left[\sum_i \left(\log\frac{1-k_i}{1+k_i} - \log\frac{1-k^{(g)}_i}{1+k^{(g)}_i}\right)^2 \right]\ .
  \label{eq:lar-regularization}
\end{equation}

\subsubsection{Log-area ratio matching}
\label{sec:LAR-matching}
As a last option, we consider using only~(\ref{eq:lar-regularization}) to adapt
the LPC estimation, while using the standard discrete
cross-entropy (non-compensated) from~(\ref{eq:loss-cross-entropy}). In that scenario, the LPCs
attempt to match the ground truth LPCs without any regard for the output probability
estimation process. The excitation is still computed based on the predicted LP coefficients.

\section{Evaluation}
\label{sec:evaluation}

We evaluate the end-to-end LPCNet on a speaker-independent,
language-independent synthesis task where the inputs features are computed
from a reference speech signal. We train all models on 205~hours
of 16-kHz speech data from a combination of TTS datasets~\cite{demirsahin-etal-2020-open,
kjartansson-etal-2020-open,kjartansson-etal-tts-sltu2018,
guevara-rukoz-etal-2020-crowdsourcing,he-etal-2020-open,oo-etal-2020-burmese,
van-niekerk-etal-2017,gutkin-et-al-yoruba2020,bakhturina2021hi}. The data includes
more than 900~speakers in 34~languages and dialects.

The training procedure includes noise augmentation in a similar way to the one
described in~\cite{valin2019lpcnetcodec}. To make the synthesis more robust to
different microphone impulse responses, we include
random spectral augmentation filtering~\cite{valin2018rnnoise}.
Training sequences span 150~ms (using 10-ms frames), with 128~sequences per batch. 
All models are trained for 20~epochs (767k~updates). For the end-to-end
models, we add an extra epoch at the end of training, where the frame rate network
weights are frozen so that the sample rate network can learn based on the final LPCs. 

We evaluate the models on a combination of the PTDB-TUG speech corpus~\cite{PirkerWPP11} and the
NTT Multi-Lingual Speech Database for Telephonometry. From PTDB-TUG, we use
all English speakers (10~male, 10~female) and randomly pick 200~concatenated pairs
of sentences. For the NTT database, we select the American English and British
English speakers (8~male, 8~female), which account for a total of 192~samples
(12~samples per speaker). The training material does not include any speakers from
the evaluation datasets. 

We compare the original LPCNet with the three end-to-end variants from 
Sections~\ref{sec:L1-regularization}, \ref{sec:LAR-regularization},
and~\ref{sec:LAR-matching}, denoted as $L_1$, LAR, and LAR/CE, respectively.
For all models, we test with block sparse $\mathrm{GRU_A}$ sizes of 192, 384, and 640~units,
using weight densities of 0.25, 0.1, and 0.15, respectively. The size of $\mathrm{GRU_B}$
is set to 32~units.

In addition to the original and end-to-end LPCNet
models, we also evaluate the reference speech as an upper bound on quality,
and we include the Speex 4~kb/s wideband vocoder~\cite{valin2007speex} as a low anchor.

\section{Results}
\label{sec:results}

\begin{table}[t]
\caption{Log-spectral distance (LSD) between the end-to-end estimated
LPCs and the ground truth LPCs over active frames of the evaluation set.\label{tab:LSD}}
\centering{}
\begin{tabular}{lccc}
\hline
Model & \multicolumn{3}{c}{LSD (dB)} \\
\cline{2-4}
$\mathrm{GRU_A}$ units & 192 & 384 & 640 \\
\hline
End-to-end $L_1$  & 3.58 & 3.64 & 3.64 \\
End-to-end LAR    & 0.34 & 0.32 & 0.46 \\
End-to-end LAR/CE & 0.87 & 0.86 & 1.07 \\
\hline
\end{tabular}
\end{table}

\begin{figure}[t]
\centering
\includegraphics[width=1.0\columnwidth]{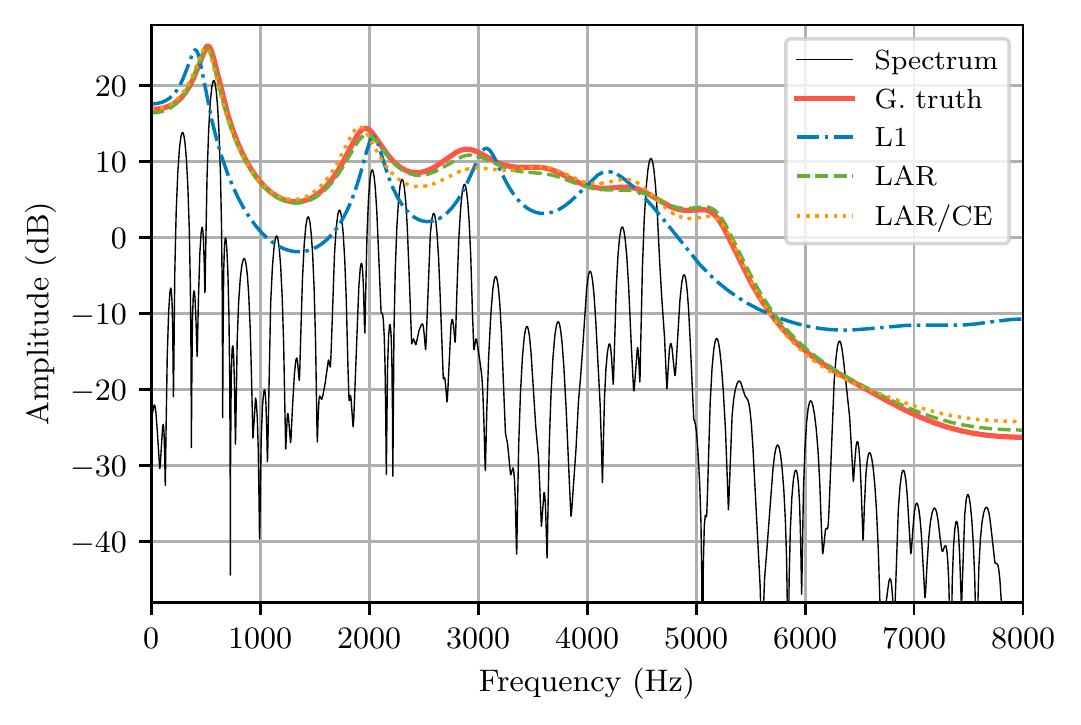}
\caption{Example spectrum for a pre-emphasized voiced frame of a female speaker. We
show the LPC responses obtained with all models of size~384. The LAR
and LAR/CE models follow the ground truth LPCs closely
(LSD of 0.53~dB and 1.12~dB, respectively), but the $L_1$-regularized
model has slightly different resonances and a higher "floor" (LSD of 7.17~dB).}
\label{fig:lpc_response}
\end{figure}

We first compare the LPC estimation behavior of the different end-to-end variants.
More specifically, we compare the response of the estimated LPCs with that of the
explicit LPC analysis. Even though we refer to the latter as the \textit{ground truth},
there is no assumption that it necessarily results in better synthesis quality than
other LPC estimation methods. We measure the log-spectral distance between the
ground truth LPCs and the end-to-end models on the active frames of all
384~test samples. The results in Table~\ref{tab:LSD} show that both LAR-regularized
models match the ground truth LPCs closely, whereas the $L_1$-regularized
models show a moderate deviation from the ground truth LPCs.
This is not unexpected,
since the training does not attempt to match the ground truth explicitly, but rather
minimize the residual.
Fig.~\ref{fig:lpc_response}
provides an example of LPC responses for the different models. 

\begin{figure}[t]
\centering
\includegraphics[width=0.85\columnwidth]{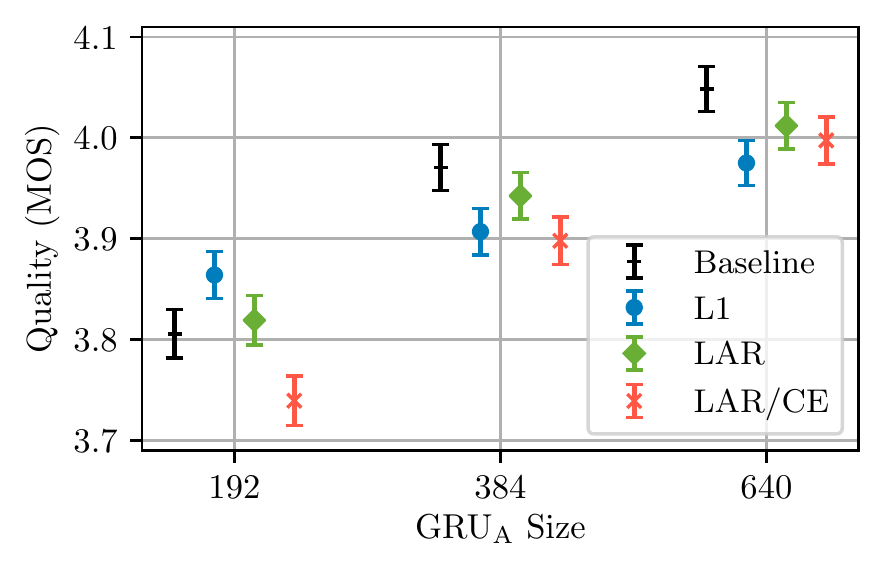}
\caption{Results from the MOS quality evaluation, including the 95\% confidence intervals.
The reference speech has a MOS of $4.21\pm0.02$ and the Speex anchor has a MOS
of~$2.76\pm0.04$.\label{fig:Quality-MOS}}
\end{figure}

We evaluate quality using mean opinion score (MOS)~\cite{P.800} using the
crowdsourcing methodology described in P.808~\cite{P.808}. Each speech sample is evaluated
by 20~randomly-selected listeners. The results in Fig.~\ref{fig:Quality-MOS}
show that both the $L_1$- and LAR-regularized models perform similarly to the
original LPCNet model, although they are significantly worse ($p<.05$) on some sizes.
On the other hand, the LAR/CE model performs significantly worse than the baseline at all sizes
and is also significantly worse than LAR on 2~sizes. That demonstrates the usefulness of
the compensated, interpolated cross-entropy loss in the end-to-end training process.
Further, our results demonstrate the our model’s output quality is not directly linked to
how closely the estimated LPCs match the ground truth LPCs.

\begin{figure}[t]
\centering
\includegraphics[width=0.85\columnwidth]{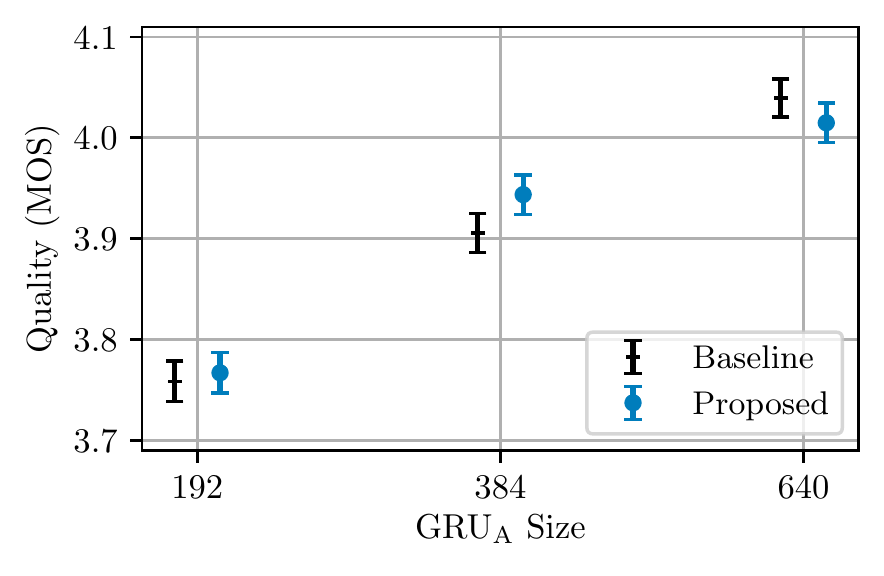}
\caption{Results from the MOS quality evaluation, including the 95\% confidence intervals.
The reference speech has a MOS of $4.21\pm0.02$ and the Speex anchor has a MOS
of~$2.56\pm0.03$. It is clear that the proposed method performs as well or
better than the baseline, while providing the extra advantage of being fully-differentiable.
\label{fig:Quality-MOS-final}}
\end{figure}

To close the slight quality gap found in the results of Fig.~\ref{fig:Quality-MOS}, we
propose to use both $L_1$ and LAR regularization. We evaluate the new samples using the same
methodology, with 20~randomly-selected listeners per sample. The results in
Fig.~\ref{fig:Quality-MOS-final} show that the proposed model is significantly better than
the baseline model at size~384 and is statistically tied for the other two sizes.

\section{Conclusion}
\label{sec:conclusion}

We have proposed an end-to-end version of LPCNet where the LPCs no
longer need to be explicitly computed. We also demonstrated a compensated,
interpolated cross-entropy
loss that improves the performance of the trained end-to-end models.
The end-to-end LPCNet achieves
equal or better synthesis quality compared to the original LPCNet, while removing the
constraint of having interpretable acoustic features for the conditioning.
The resulting vocoder architecture makes it easier to use LPCNet, for a broader
range of applications.
We believe that this work can be extended to bring the benefits of
linear prediction to other ML-based audio tasks, such as time-domain
speech enhancement systems.
It is an open question whether the proposed technique could also be
applied to non-autoregressive neural synthesis techniques, such as GAN-
and flow-based models.

\balance
\bibliographystyle{IEEEbib}
\bibliography{refs,corpora}

\begin{thebibliography}{10}

\bibitem{markel1974linear}
J.~Markel and A.~Gray,
\newblock ``A linear prediction vocoder simulation based upon the
  autocorrelation method,''
\newblock {\em IEEE Transactions on Acoustics, Speech, and Signal Processing},
  vol. 22, no. 2, pp. 124--134, 1974.

\bibitem{hunt1996unit}
A.J. Hunt and A.W. Black,
\newblock ``Unit selection in a concatenative speech synthesis system using a
  large speech database,''
\newblock in {\em Proc. International Conference on Acoustics, Speech and
  Signal Processing (ICASSP)}, 1996, vol.~1, pp. 373--376.

\bibitem{oord2016wavenet}
A.~van~den Oord, S.~Dieleman, H.~Zen, K.~Simonyan, O.~Vinyals, A.~Graves,
  N.~Kalchbrenner, A.~Senior, and K.~Kavukcuoglu,
\newblock ``{WaveNet}: A generative model for raw audio,''
\newblock {\em arXiv preprint arXiv:1609.03499}, 2016.

\bibitem{kalchbrenner2018efficient}
N.~Kalchbrenner, E.~Elsen, K.~Simonyan, S.~Noury, N.~Casagrande, E.~Lockhart,
  F.~Stimberg, A.~van~den Oord, S.~Dieleman, and K.~Kavukcuoglu,
\newblock ``Efficient neural audio synthesis,''
\newblock in {\em Proc. International Conference on Machine Learning (ICML)},
  2018, pp. 2410--2419.

\bibitem{donahue2019wavegan}
C.~Donahue, J.~McAuley, and M.~Puckette,
\newblock ``Adversarial audio synthesis,''
\newblock in {\em Proc. ICLR}, 2019.

\bibitem{prenger2019waveglow}
R.~Prenger, R.~Valle, and B.~Catanzaro,
\newblock ``{WaveGlow}: A flow-based generative network for speech synthesis,''
\newblock in {\em Proc. International Conference on Acoustics, Speech and
  Signal Processing (ICASSP)}, 2019.

\bibitem{valin2019lpcnet}
J.-M. Valin and J.~Skoglund,
\newblock ``{LPCNet}: Improving neural speech synthesis through linear
  prediction,''
\newblock in {\em Proc. International Conference on Acoustics, Speech and
  Signal Processing (ICASSP)}. IEEE, 2019, pp. 5891--5895.

\bibitem{fant1960acoustic}
G.~Fant,
\newblock {\em Acoustic theory of speech production},
\newblock Walter de Gruyter, 1960.

\bibitem{valin2019lpcnetcodec}
J.-M. Valin and J.~Skoglund,
\newblock ``A real-time wideband neural vocoder at 1.6 kb/s using {LPCNet},''
\newblock in {\em Proc. INTERSPEECH}, 2019.

\bibitem{kons2019tts}
Z.~Kons, S.~Shechtman, A.~Sorin, C.~Rabinovitz, and R.~Hoory,
\newblock ``High quality, lightweight and adaptable {TTS} using {LPCNet},''
\newblock in {\em Proc. INTERSPEECH}, 2019.

\bibitem{zhao2021conversion}
S.~Zhao, H.~Wang, T.H. Nguyen, and B.~Ma,
\newblock ``Towards natural and controllable cross-lingual voice conversion
  based on neural {TTS} model and phonetic posteriorgram,''
\newblock in {\em Proc. International Conference on Acoustics, Speech and
  Signal Processing (ICASSP)}, 2021.

\bibitem{zeghidour2021soundstream}
N.~Zeghidour, A.~Luebs, A.~Omran, J.~Skoglund, and M.~Tagliasacchi,
\newblock ``{SoundStream}: An end-to-end neural audio codec,''
\newblock 2021.

\bibitem{donahue2021tts}
J.~Donahue, S.~Dieleman, M.~Bińkowski, E.~Elsen, and K.~Simonyan,
\newblock ``End-to-end adversarial text-to-speech,''
\newblock in {\em Proc. International Conference on Learning Representations
  (ICLR)}, 2021.

\bibitem{gupta2019bwe}
A.~Gupta, B.~Shillingford, Y.~Assael, and T.C. Walters,
\newblock ``Speech bandwidth extension with {WaveNet},''
\newblock in {\em Proc. IEEE Workshop on Applications of Signal Processing to
  Audio and Acoustics (WASPAA)}, 2019.

\bibitem{makhoul1975linear}
J.~Makhoul,
\newblock ``Linear prediction: A tutorial review,''
\newblock {\em Proceedings of the IEEE}, vol. 63, no. 4, pp. 561--580, 1975.

\bibitem{itakura1975line}
F.~Itakura,
\newblock ``Line spectrum representation of linear predictive coefficients of
  speech signals,''
\newblock {\em Journal of the Acoustical Society of America}, vol. 57, no. S1,
  1975.

\bibitem{viswanathan1975quant}
R.~Viswanathan and J.~Makhoul,
\newblock ``Quantization properties of transmission parameters in linear
  predictive systems,''
\newblock {\em IEEE Transactions on Acoustics, Speech, and Signal Processing},
  vol. 23, no. 3, pp. 309--321, 1975.

\bibitem{levinson1947lpc}
N.~Levinson,
\newblock ``The {Wiener} {RMS} error criterion in filter design and
  prediction,''
\newblock {\em Journal of Mathematical Physics}, vol. 25, pp. 261--–278,
  1947.

\bibitem{rao2020sfnet}
A.~Rao and P.K. Ghosh,
\newblock ``{SFNet}: A computationally efficient source filter model based
  neural speech synthesis,''
\newblock {\em IEEE Signal Processing Letters}, vol. 27, pp. 1170--1174, 2020.

\bibitem{demirsahin-etal-2020-open}
I.~Demirsahin, O.~Kjartansson, A.~Gutkin, and C.~Rivera,
\newblock ``{Open-source Multi-speaker Corpora of the English Accents in the
  British Isles},''
\newblock in {\em Proc. LREC}, 2020.

\bibitem{kjartansson-etal-2020-open}
O.~Kjartansson, A.~Gutkin, A.~Butryna, I.~Demirsahin, and C.~Rivera,
\newblock ``{Open-Source High Quality Speech Datasets for Basque, Catalan and
  Galician},''
\newblock in {\em Proc. SLTU and CCURL}, 2020.

\bibitem{kjartansson-etal-tts-sltu2018}
K.~Sodimana, K.~Pipatsrisawat, L.~Ha, M.~Jansche, O.~Kjartansson, P.~De Silva,
  and S.~Sarin,
\newblock ``{A Step-by-Step Process for Building TTS Voices Using Open Source
  Data and Framework for Bangla, Javanese, Khmer, Nepali, Sinhala, and
  Sundanese},''
\newblock in {\em Proc. SLTU}, 2018.

\bibitem{guevara-rukoz-etal-2020-crowdsourcing}
A.~Guevara-Rukoz, I.~Demirsahin, F.~He, S.-H.~C. Chu, S.~Sarin,
  K.~Pipatsrisawat, A.~Gutkin, A.~Butryna, and O.~Kjartansson,
\newblock ``{Crowdsourcing Latin American Spanish for Low-Resource
  Text-to-Speech},''
\newblock in {\em Proc. LREC}, 2020.

\bibitem{he-etal-2020-open}
F.~He, S.-H.~C. Chu, O.~Kjartansson, C.~Rivera, A.~Katanova, A.~Gutkin,
  I.~Demirsahin, C.~Johny, M.~Jansche, S.~Sarin, and K.~Pipatsrisawat,
\newblock ``{Open-source Multi-speaker Speech Corpora for Building Gujarati,
  Kannada, Malayalam, Marathi, Tamil and Telugu Speech Synthesis Systems},''
\newblock in {\em Proc. LREC}, 2020.

\bibitem{oo-etal-2020-burmese}
Y.~M. Oo, T.~Wattanavekin, C.~Li, P.~De~Silva, S.~Sarin, K.~Pipatsrisawat,
  M.~Jansche, O.~Kjartansson, and A.~Gutkin,
\newblock ``{Burmese Speech Corpus, Finite-State Text Normalization and
  Pronunciation Grammars with an Application to Text-to-Speech},''
\newblock in {\em Proc. LREC}, 2020.

\bibitem{van-niekerk-etal-2017}
D.~van Niekerk, C.~van Heerden, M.~Davel, N.~Kleynhans, O.~Kjartansson,
  M.~Jansche, and L.~Ha,
\newblock ``{Rapid development of TTS corpora for four South African
  languages},''
\newblock in {\em Proc. Interspeech}, 2017.

\bibitem{gutkin-et-al-yoruba2020}
A.~Gutkin, I.~Demir{\c{s}}ahin, O.~Kjartansson, C.~Rivera, and
  K.~T{\'u}b\d{\`o}s{\'u}n,
\newblock ``{Developing an Open-Source Corpus of Yoruba Speech},''
\newblock in {\em Proc. INTERSPEECH}, 2020.

\bibitem{bakhturina2021hi}
E.~Bakhturina, V.~Lavrukhin, B.~Ginsburg, and Y.~Zhang,
\newblock ``{Hi-Fi Multi-Speaker English TTS Dataset},''
\newblock {\em arXiv preprint arXiv:2104.01497}, 2021.

\bibitem{valin2018rnnoise}
J.-M. Valin,
\newblock ``A hybrid {DSP}/deep learning approach to real-time full-band speech
  enhancement,''
\newblock in {\em Proceedings of IEEE Multimedia Signal Processing (MMSP)
  Workshop}, 2018.

\bibitem{PirkerWPP11}
G.~Pirker, M.~Wohlmayr, S.~Petrik, and F.~Pernkopf,
\newblock ``A pitch tracking corpus with evaluation on multipitch tracking
  scenario.,''
\newblock in {\em Proc. INTERSPEECH}, 2011, pp. 1509--1512.

\bibitem{valin2007speex}
J.-M. Valin,
\newblock ``The {Speex} codec manual,'' 2007.

\bibitem{P.800}
ITU-T,
\newblock {\em Recommendation {P}.800: Methods for subjective determination of
  transmission quality}, 1996.

\bibitem{P.808}
ITU-T,
\newblock {\em Recommendation {P}.808: Subjective evaluation of speech quality
  with a crowdsourcing approach}, 2018.

\end{thebibliography}

\end{document}